\documentclass[aps,preprint,showpacs,superscriptaddress]{revtex4-1}  
\usepackage{graphicx}
\usepackage{subcaption}
\usepackage{bm}   
\usepackage{color}
\usepackage{amssymb}
\usepackage{amsmath}
\usepackage{float}

\begin{document}
	
	\title[Symmetrical emergence of extreme events]{Symmetrical emergence of extreme events at multiple regions in a damped and driven velocity-dependent mechanical system}
	
	\author{S. Sudharsan}
	\affiliation{Department of Nonlinear Dynamics, Bharathidasan University, Tiruchirappalli - 620 024, Tamilnadu, India}
	\author{A. Venkatesan}
	\affiliation{PG \& Research Department of Physics, Nehru Memorial College (Autonomous), Puthanampatti, Tiruchirappalli - 621 007,  Tamil Nadu, India.}
	\author{M. Senthilvelan}
	\email[Correspondence to: ]{velan@cnld.bdu.ac.in}
	\affiliation{Department of Nonlinear Dynamics, Bharathidasan University, Tiruchirappalli - 620 024, Tamilnadu, India}
	\vspace{10pt}
	
	\begin{abstract}
		In this work, we report the emergence of extreme events in a damped and driven velocity-dependent mechanical system. We observe that the extreme events emerge at multiple points. We further notice that the extreme events occur symmetrically in both positive and negative values  at all the points of emergence. We statistically confirm the emergence of extreme events by plotting the probability distribution function of peaks and interevent intervals. We also determine the mechanism behind the emergence of extreme events at all the points and classify these points into two categories depending on the region at which the extreme events emerge. Finally, we plot the two parameter diagram in order to have a complete overview of the system.  
	\end{abstract}
	
	%
	%
	%
	%
	%
	\maketitle
	\section{Introduction}
	Events that occur suddenly and intermittently in natural systems, man-made systems and societal systems are classified as extreme events. For example, rogue waves, floods, cyclones, tsunamies, tornadoes, earthquakes, droughts, epidemics, epileptic seizures are extreme events that occur in natural systems, while material ruptures, explosions, nuclear disasters, chemical contaminations are extreme events happening in man-made systems whereas financial crisis, share market crashes, ecological regime shifts are extreme events which come across in societal systems \cite{Krause2015, Jentsch2005, threshold}.  Due to its catastrophic nature, predicting the extreme events is much needed in order to avoid or mitigate enormous destruction to the nature and to the society. One way to predict and control these events is to study the dynamics underlying the onset of these events.
	
	\par Four important aspects involved in the dynamical study of extreme events are (i) mechanisms, (ii) prediction, (iii) mitigation and (iv) statistics \cite{extremereview}. Mechanism involves understanding the way in which  these events are being generated. Prediction involves determining the measurable observables called reliable indicators which act as an indicator of extreme events in the case of real-time prediction. Mitigation is determining the ways to control the emergence of extreme events from a dynamical systems framework. Finally, statistics helps to find out the frequency and probability in which extreme events will reoccur in a given circumstance. To study the last two aspects, initially it is important to predict the occurrence and to determine the mechanism underlying the occurrence of extreme events.
	
	\par As far as dynamical systems are concerned, extreme events are observed in FitzHugh - Nagumo oscillators,  Hindmarsh - Rose model, populations dynamics, Li\'enard system, micro - mechanical system, electronic circuits, coupled Ikeda map, network of Josephson junctions and nonlinear Schr\"odinger equation \cite{feudal1, feudal2, hr, ricker, leo, suresh, circuit, ikeda, dana, nls}. In regard of applications, these events have been found to occur in laser systems, plasma, optical fibers etc. \cite{laser1, plasma, optfib2}. Experimentally, it has been found in epileptic EEG studies in rodents, annular wave flume and climatic studies \cite{eeg, flume, kurth1}.
	
	\par As mentioned earlier, emergence and determination of the mechanisms are important tasks in the study of extreme events. There are several mechanisms proposed, in the literature, as the precursors. In the literature, the emergence of extreme events were sudied by investigating the asymptotic behaviour of the system usually termed as attractor. In most of the studies, it was observed that the extreme events occur due to sudden change or expansion of irregular and unpredictable asymptotic behaviour of the system or simply chaotic attractor. In particular, in Refs. \cite{feudal1, feudal2}, extreme events were observed to emerge after an interior crisis \cite{feudal1, feudal2, leo, ikeda, co2} owing to the opening of a channel like structure in the phase space whereas in Ref. \cite{ikeda} it has been noted that the extreme events emerge at the point of interior crisis where the already existing chaotic attractor had undergone an intermittent expansion. The difference between a standard chaotic attractor and a chaotic attractor which produces extreme events is that, in a standard chaotic attractor, when the the probabilities of the peak values are distributed it will be of Gaussian type. Whereas for the chaotic attractor that exhibits extreme events, the peak distribution is long tailed. In a mechanical Li\'enard system \cite{leo}, extreme events were found to emerge at two places, (i) at the point of interior crisis and (ii) at the point where periodic oscillation takes an intermittency route to chaos. It has also been pointed out that the collision of a chaotic attractor with a saddle orbit as the reason behind the emergence of extreme events \cite{leo}. In addition to this, in Ref. \cite{suresh}, sticking and sliding of the system's trajectory near the discontinuous boundary were proposed as a reason behind the emergence of extreme events. The other possible reasons for the occurrence of extreme events as given in the literature are (i) transient instabilities \cite{royal}, (ii) influence of noise in multistable systems \cite{multi} and (iii) attractor bubbling \cite{atbub}.  Despite having found a few mechanisms, exploring other unknown mechanisms for larger and more general classes of systems are underway. By identifying and studying the precursors behind the origin of extreme events, we can determine the various factors which can control its occurrence.
	
	\par Since there are no formal definitions for extreme events, dynamically extreme events are said to be an occasional, rare but recurrent excursion by an observable to a larger value that exceeds an abnormality index, amid the regular small-scale fluctuations \cite{threshold}. Statistically, this abnormality index is defined as the average height of all peaks in a measured time series plus 4 (or greater than 4) times the standard deviation of the peaks. Further, a fat tailed behaviour of the distributed peaks and exponential distribution of the inter-event interval  (due to the Poissonian process) are some of the important properties which are required to identify an event to be extreme \cite{peakpdf1}.
	
	\par In this work, we report the emergence of extreme events and determine the mechanism by which extreme events occur in a damped and driven velocity-dependent mechanical system. For this, we consider a well known mechanical model \cite{nay} which is known for exhibiting rich nonlinear dynamics \cite{av}. Typically this model represents the dynamics of a particle in a rotating parabola. This mechanical model also describes a motor bike being rided in a rotating parabolic well in a circus, centrifugation devices, centrifugal filters and industrial hoppers \cite{av, centrifugation, filter, hopper}. Extreme events in these systems may represent a sudden increase in the velocity of the particle inside the rotating system which may cause sudden damage to the system. Detecting extreme events in such a system will definitely help us to mitigate unexpected accident to the particle and also to the system themselves. Since the system which we consider is a mechanical one, detecting extreme events in this system can also help us to  redesign the engineering so that extreme events can be avoided. 
	
	\par  Our results reveal that extreme events emerge at multiple regions in the considered mechanical system. In all the previous studies made in single systems on the emergence of extreme events, there would be an interior crisis occurring which in turn lead to the expansion of chaotic attractor and this expanded chaotic attractor would either had continued to sustain or had destroyed by boundary crisis \cite{leo}. In other words extreme events tend to appear at the beginning and end of a single chaotic attractor \cite{leo}. In our work, we observe that such a mechanism is occuring at multiple regions and hence we call it as emergence of extreme events at multiple regions. Here emergence explicitly represents the initiation of extreme events. Such multiple initiations have already been implicitly reported in coupled systems \cite{feudal2, ikeda} and in the loss-modulated $CO_2$ laser model \cite{co2}. Differing from this, for the first time, we observe a similar behaviour in a mechanical system. In addition to this, in our system, we observe the emergence of symmetrical negative valued and positive valued extreme events. To our knowledge, this is the first time a mechanical system which can generate extreme events symmetrically in both spatial directions is being reported. Also it is the first time, the emergence of extreme events in a non-polynomial mechanical oscillator/velocity-dependent potential system is being studied.

	\par We organize our work as follows. In Sec. 2, we investigate and study the bifurcation of a non-polynomial velocity-dependent system. In Sec. 3, we report the emergence of extreme events in both positive and negative spatial directions at multiple regions. In Sec. 4, we analyze the mechanism underlying these extreme events which occur at multiple regions. Finally, in Sec. 5, we summarize our work.
	
	\section{Model}
	We consider a mechanical model which describes the motion of a freely sliding particle of unit mass
	on a parabolic wire rotating with a constant angular velocity $\omega_0$ about the axis $z=\sqrt{\lambda}x^2$. Here $1/\sqrt{\lambda}$ is the semi-latus rectum of the rotating parabola and $\lambda,~\omega_0>0$ \cite{nay}. The associated equation of motion can be formulated in the form
	\begin{equation}
	(1+\lambda x^2)\ddot{x}+\lambda x \dot{x}^2 +\omega_0^2x=0,
	\label{first}
	\end{equation}
	where overdot represents time derivative. Subjecting Eq. (\ref{first}) to additional linear damping and periodic forcing, we get
	\begin{equation}
	(1+\lambda x^2)\ddot{x}+\lambda x \dot{x}^2 +\omega_0^2x+\alpha \dot{x}=f~\mathrm{cos}\omega t,
	\label{master}
	\end{equation}
	where $\lambda$, $\omega_0$ and $\alpha$ represents the same parameters as described above, while $f$ is the external forcing strength and $\omega$ is the frequency of the external drive.
	
	For numerical integration purpose, we rewrite Eq. (\ref{master}) as
	
	\begin{eqnarray}
	\dot{x}&=&y,\nonumber\\
	\dot{y}&=&\frac{f \cos  \omega_e t -\alpha y-\lambda x y^2 - \omega_0^2 x}{1 + \lambda x^2}.
	\label{couppardri}
	\end{eqnarray} 
	
	The numerical integration was carried out using fourth order Runge Kutta method. 
	In the literature, a variety of nonlinear dynamical analysis have been made on (\ref{first}) and (\ref{master}) in an extensive manner. To name a few we cite the following. System (\ref{master}) is well known for its exhibition of rich bifurcation scenarios. In particular, (\ref{master}) has been analysed in \cite{av} and was found to exhibit several bifurcation scenarios such as symmetry breaking, period doubling, periodic windows, intermittency and antimonotonicity. Further, the quasiperiodically driven version of (\ref{first}) has also been studied thoroughly in Ref. \cite{av} and it is found to exhibit two routes to chaos, namely (i) torus doubling, torus merging, wrinkling, strange non-chaotic attractor (SNA), and chaos and (ii) two-frequency quasiperiodicity, torus doubling, wrinkling, SNA, chaos, SNA, wrinkling, inverse torus doubling, torus, torus bubbles followed by the onset of torus breaking to chaos via SNA or followed by the onset of torus doubling route to chaos via SNAs \cite{av1}. The complete chaotic dynamics of a slightly different version of Eq. (\ref{first}), well known as Mathews-Lakshmanan oscillator, endowed with position dependent mass along with damping and external forcing has been studied in \cite{ghosh}. 
	
	\begin{figure}[!ht]
		\centering
		\includegraphics[width=8.6cm]{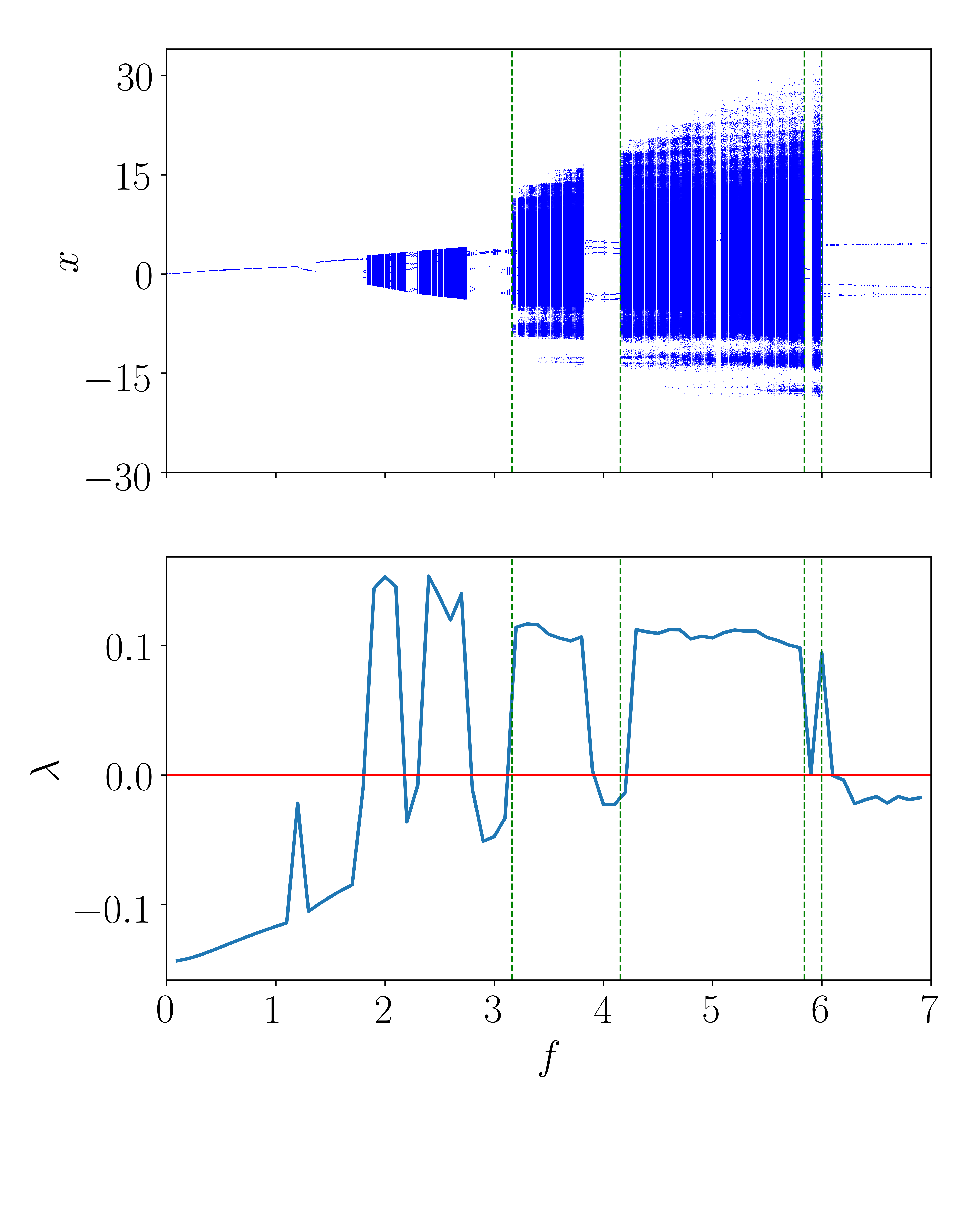}
		\caption{(a) Bifurcation plot of (\ref{couppardri}) and (b) Lyapunov exponent for varying $f$. Other parameter values are $\omega_0^2=0.25$, $\lambda=0.5$, $\alpha=0.2$, $\omega=1.0$. The green dotted vertical lines at $f=3.1608$, $f=4.15601$, $f=5.84001$ and $f=5.998$ represents the parameter values for which extreme events originate.} 
		\label{bifurcate}
	\end{figure}
	
	In the present work, we intend to analyse the emergence of extreme events in (\ref{master}). We fix the parameter values as $\omega_0^2=0.25$, $\lambda=0.5$, $\alpha=0.2$, $\omega=1.0$ just as in Ref. \cite{av}. The bifurcation plot of Eq. (\ref{master}) \cite{av} is shown in Fig. \ref{bifurcate}. The nonlinear dynamics of Eq. (\ref{master}) has been studied for the parameter domain ranging from $f=0.1$ to $3.16$ \cite{av}. In our present study, we concentrate on the parameter region beyond $f=3.16$ where we observe generation of extreme events at multiple regions. The initial conditions are choosen as $x(0)=\dot{x}(0)=-0.01$ and fixed throughout the work. As far as Eq. (\ref{master}) is concerned, increasing the external forcing strength beyond $f=3.0$ period doubling scenario occurs. At $f=3.1608$ interior crisis (here interior crisis represents sudden expansion of chaotic attractor) occurs leading to the sudden expansion of the chaotic attractor. After such a sudden expansion, the amplitude of the attractor continues to grow gradually. Meanwhile, for a small parameter range between $f=3.1934$ and $f=3.2124$ reverse period doubling and period doubling occurs. At $f=3.8249$ tangent bifurcation occurs and the expanded chaotic attractor suddenly turns into a $5T$-periodic and sustains till $f=4.15601$ where suddenly an expanded chaotic attractor appears. Now again the amplitude of the chaotic attractor continues to expand gradually till $f=5.99865$ after which the interior crisis occurs (in a reverse manner) and the chaotic orbit gets destroyed to become a periodic orbit. Even during the expansion, periodic window appears due to the tangent bifurcation between $f=5.84001$ and $f=5.92000$. The transition from periodic to chaotic nature can be clearly seen from the transition of the  Lyapunov exponent which changes its sign from negative to positive as shown in Fig.~\ref{bifurcate}(b). 
	
	\section{Observation of Extreme events}
	\label{sec:3}
	In the literature, events are classified as extreme events if the trajectory of the system seldom crosses a particular threshold value \cite{threshold}. The threshold value is fixed using the relation $x_{ee}=\langle x \rangle + n\sigma_x$, where $\langle x \rangle$ is the mean peak amplitude, $\sigma_x=\big(\langle x^2 \rangle - \langle x \rangle^2\big)^{1/2}$ is the standard deviation over mean peak amplitude and $4\leq n \leq 8$ \cite{ikeda}. The crossing of the system's trajectory over this  threshold value can occur even at a very large time. Throughout this work we fix $n=4$ and we calculate the value of $x_{ee}$ from very long time iterations of about $10^9$ iterations. Here iteration means computing the value of state variables of the considered dynamical system using fourth order Runge Kutta method. While implementing this algorithm a step size of $0.01$ was taken. For positive threshold max$(x)$'s are considered and for negative threshold min$(x)$'s are considered. The corresponding relation for the calculation of negative threshold is $x_{ee}=\langle x\rangle$ - 4$\sigma_x$.
	
	\par
\begin{figure}[!ht]
		\centering
		\includegraphics[width=8.6cm]{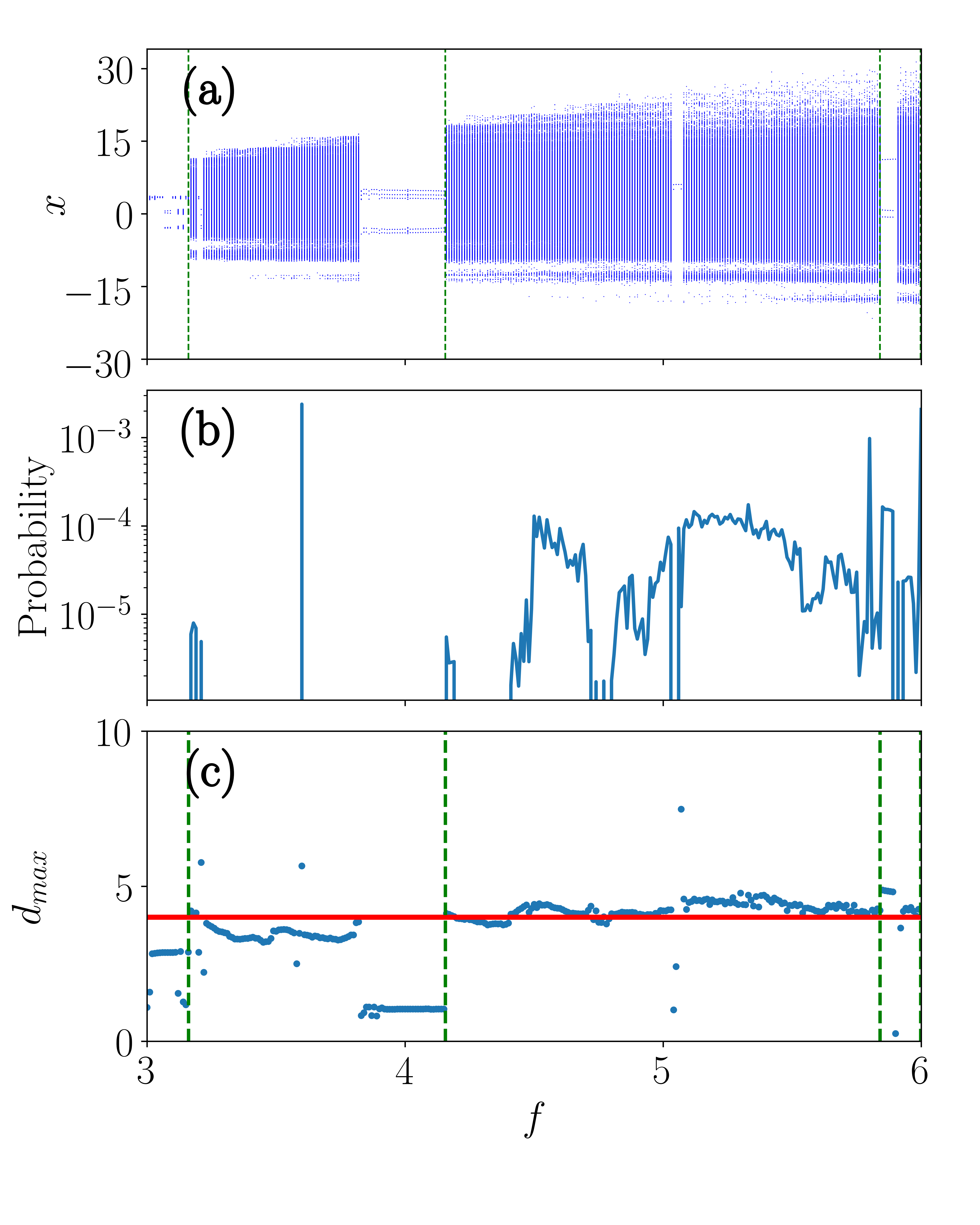}
		\caption{Plots of (a) Bifurcation plot of Eq.~(\ref{couppardri}), (b) Probability plot and (c) $d_{max}$ plot. The range of $f$ is $(3,6)$ The parameter values are $\omega_0^2=0.25$, $\lambda=0.5$, $\alpha=0.2$, $\omega=1.0$. $f=3.1608$, $f=4.15601$, $f=5.84001$ and $f=5.998$ are the parameter values for which Extreme Events originate.} 
		\label{multi}
	\end{figure}

 In the model under consideration, we observe extreme events while varying the value of the external forcing amplitude $f$. We observe the symmetrical emergence of extreme events in both positive and negative values. Upon increasing $f$ beyond $3.0$ we observe such an emergence for the following set of values of $f$, namely $f=3.1608$, $f=4.15601$, $f=5.8401$ and $f=5.998$. The point of emergence is confirmed using the enlarged bifurcation diagram, probability plot and $d_{max}$ plot which are displayed in Figs.~\ref{multi}(a), (b) and (c) respectively.
	
\begin{figure}[!ht]
		\centering
		\includegraphics[width=12cm]{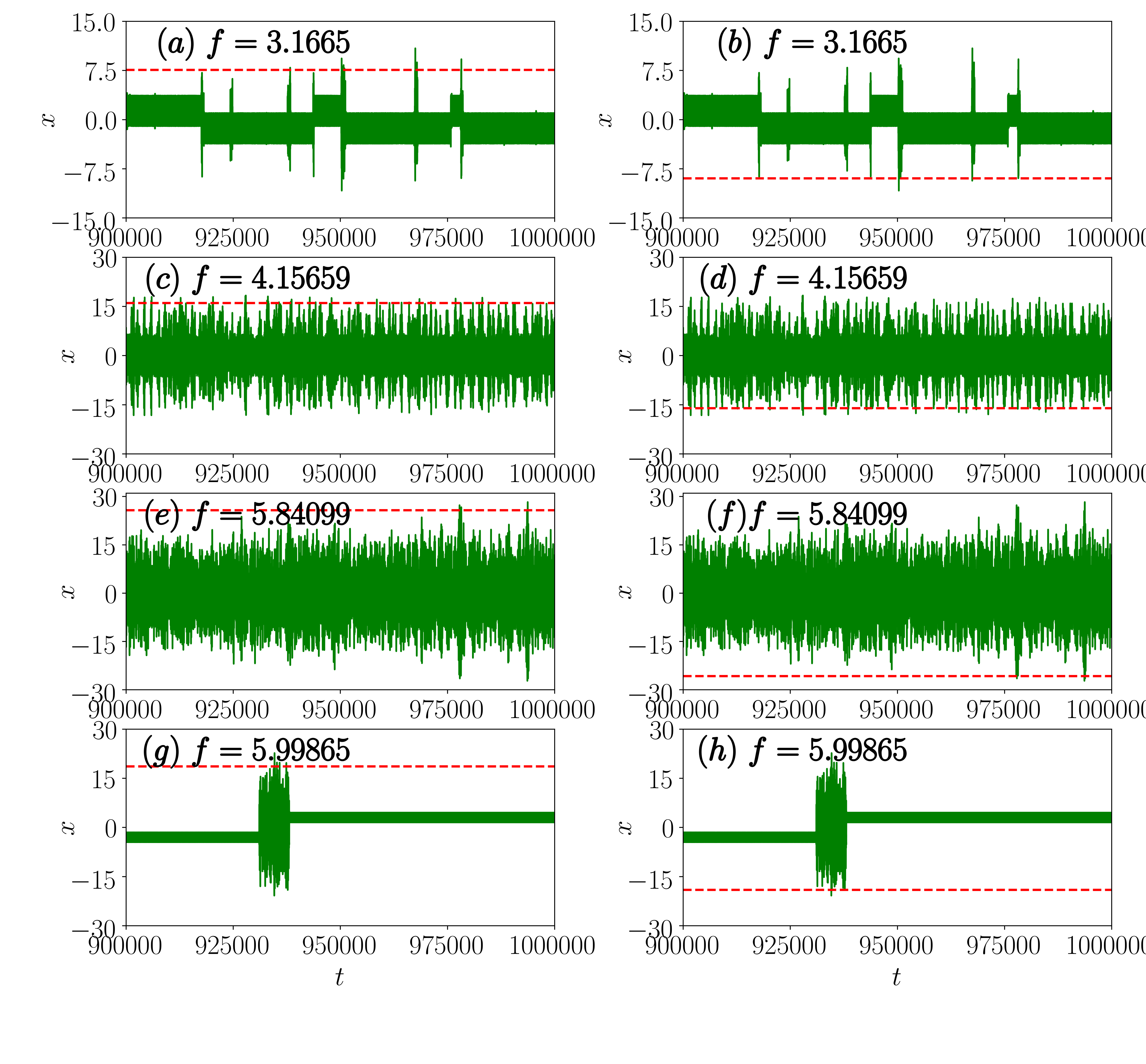}
		\caption{Time series plots of Eq. (\ref{couppardri}) showing the occurrence of symmetrical emergence of extreme events. The horizontal line represents the threshold value $x_{ee}$. Left panel represents positive valued extreme events and right panel represents negative valued extreme events. }\label{time}
	\end{figure}

	In Fig.~\ref{multi}(b), we show the probability plot for the occurrence of extreme events. We calculate the probability of occurrence of extreme events using the relation 
	\begin{equation}
	\mathrm{Probability} = \frac{\mathrm{No.~of~Peaks~crossing~ the~ threshold}}{\mathrm{Total~number~of~peaks}}.
	\end{equation}
	Regions with no spike represents regions of non-extreme events. In Fig.~\ref{multi}(c), we show the variation of $d_{max}$ against $f$. The value of $d_{max}$ is calculated using the formula 
	\begin{equation}
	d_{max}=\displaystyle\frac{\mathrm{max}(x_i)-\langle x\rangle}{\sigma_x}.
	\end{equation}
	Fig.~\ref{multi}(a) is the bifurcation plot for $f=(3,6)$. It is the enlarged version of Fig.~\ref{bifurcate} which is reproduced here for the purpose of comparison. Every time extreme events emerge, the points in the $d_{max}$ plot increases than the values in the non extreme events regime. From Fig.~\ref{multi}(b)
	we can observe that there are only few extreme events occurring in and around $f=3.1608$. Extreme events occur prominently only after $f=4.15601$. This fact can also be seen in the $d_{max}$ plot in Fig.~\ref{multi}(c).
	For the purpose of convenience in explaining the emergence of extreme events, we choose the parameter values as $f=3.1665$, $f=4.15659$, $f=5.84099$ and $f=5.99865$ in our analysis. For these parameter values, extreme events occur near the point of interior crisis when $f=3.1665$ and $f=5.99865$ whereas the extreme events occur near the point of tangent bifurcation while $f=4.15659$ and $f=5.84099$. The straight horizontal line in Fig. \ref{time} represents the calculated threshold value for each case. The time series of the system (\ref{couppardri}) which depicts the excursion of system's trajectory beyond the threshold value, both for positive and negative values of the threshold, are shown in Fig. \ref{time}. For each $f$, the first figure in column 1 in Fig. \ref{time} represents the time series with positive threshold whereas the adjacent figure in column 2 represents the time series with a negative threshold.

\begin{figure}[!ht]
		\centering
		\includegraphics[width=12cm]{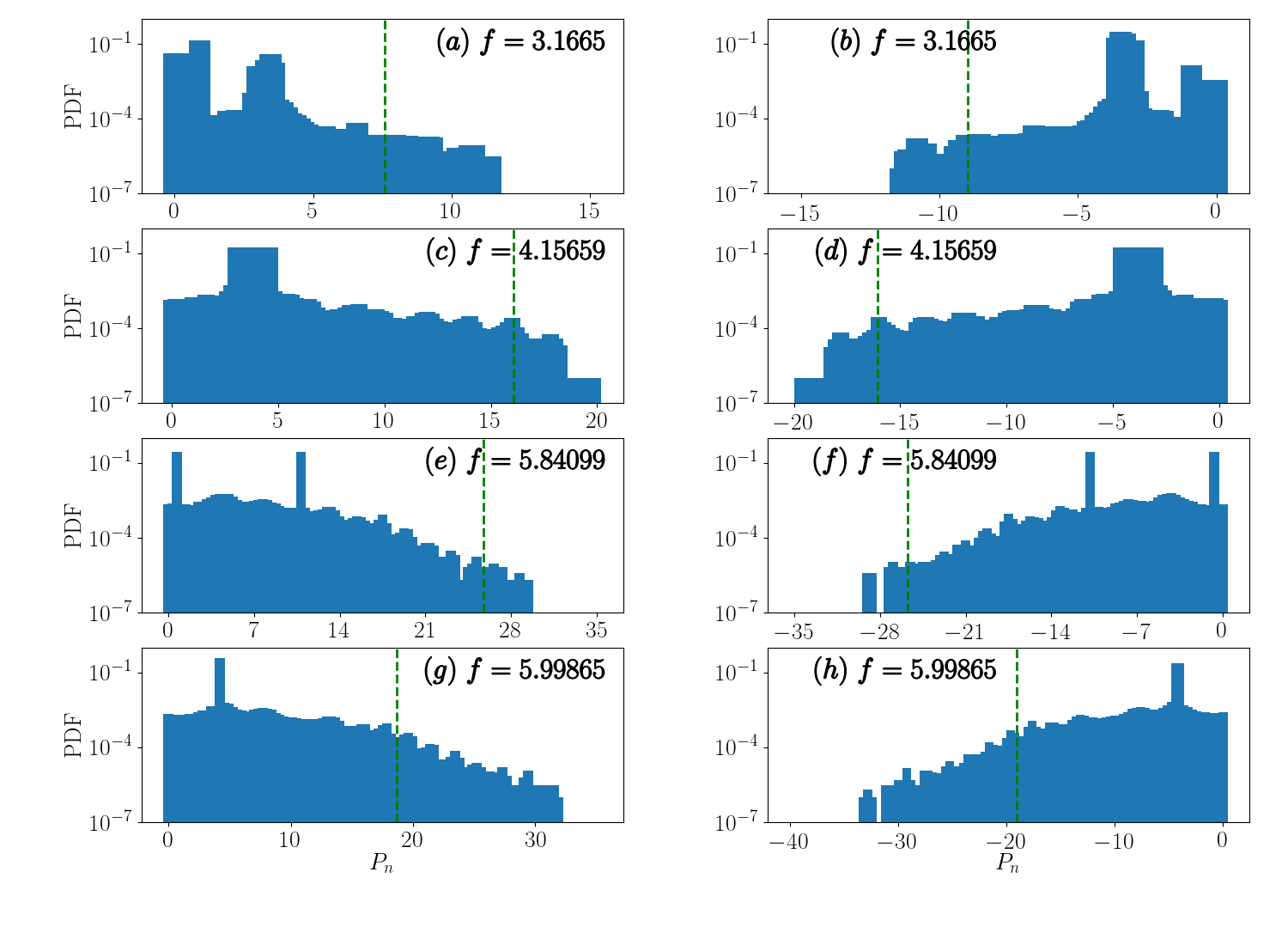}
		\caption{Peak probability density function (PDF) for different values of $f$. Left panel represents positive valued extreme events and right panel represents negative valued extreme events. The vertical line represent the threshold value $x_{ee}.$}\label{singlepdf}
	\end{figure}

	One of the important statistical characterizations of extreme events is the presence of fat-tail in the probability density function (PDF) of peaks \cite{peakpdf1} (positive peaks in the case of positive valued extreme and negative peaks in the case of negative valued extreme events). The inference that one can make from the fat-tail distribution occurring beyond the point of threshold value is that there is a small but yet significant probability for extreme events to occur. The peak PDF of all the above four cases is shown in Fig. \ref{singlepdf}.
	
	In Fig. \ref{singlepdf}, we distribute all the peaks (max ($x$) in the case of positive and min ($x$) in the case of negative) between $0$ and $1$. The vertical line in Fig. \ref{singlepdf} is the threshold $x_{ee}$ value. In all the four cases ($f=3.1665,~f=4.15659,~f=5.84099,~f=5.99865$) we can see the presence of events beyond the threshold values.  This confirms that there is atleast a small probability for extreme events to occur in the long run of the system.

\begin{figure}[!ht]
		\centering
		\includegraphics[width=12cm]{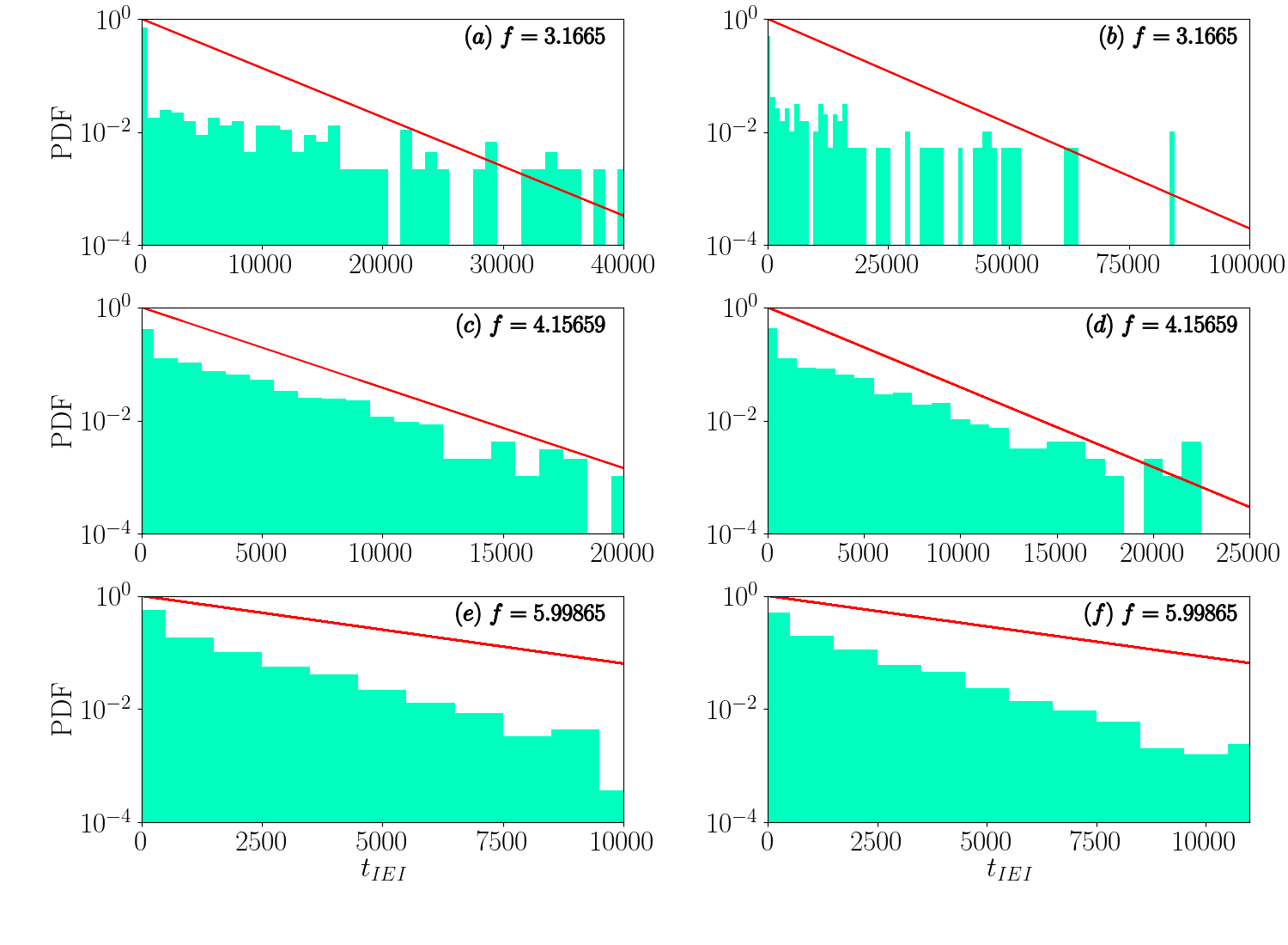}
		\caption{IEI probability density function (PDF) for different values of $f$. Left panel is for positive valued extreme events and right panel is for negative valued extreme events. The exponential fit is determined using multiples of $\mathrm{exp}(-rt_{IEI})$, where $r$ is the rate of occurrence of extreme events computed from the data and $t_{IEI}$ is the time interval between two successive extreme events.}\label{singleiei} 
	\end{figure}

	The second characterization which we have used to confirm the extreme event is the PDF of interevent intervals (IEI) \cite{feudal1}.This characterization is to confirm that extreme events are independent events. The PDF of IEI for a few parameter values, for both positive and negative spatial directions are shown in Fig. \ref{singleiei}. The $x-$axis in the plot represents time interval ($t_{IEI}$) between two successive extreme events.
	
	It is clear from Fig. \ref{singleiei} that the distribution of PDF of IEI is nearly exponential and resembling a Poissonian process confirming that the extreme events which emerge are independent events. Although the distribution of IEI for the extreme events at $f=3.1665$ differs from that of the remaining two, it still almost resembles a Poissonian distribution. At $f=5.84099$, for which we have not shown the IEI distribution here, it resembles that of $f=3.1665$.  We have also determined the rate of the occurrence, 
	\begin{equation}
	\mathrm{Rate~of~occurrence} = \frac{\mathrm{(\mathrm{Total~No. ~of ~Extreme~Events})}-1}{t_f-t_i},
	\end{equation}
	where $t_f$ is the time of the final event and $t_i$ is the time of the first extreme event for both postive and negative valued events. For positive valued events, the rate turns out that $1.99637\times10^{-4}$, $3.26565\times10^{-4}$ and $2.75128\times10^{-4}$ respectively at $f=3.1665$, $f=4.15659$ and $f=5.99865$. For negative valued events, the rates are found to be $8.48016\times10^{-5}$, $3.24498\times10^{-4}$ and $2.48295\times10^{-4}$ respectively at $f=3.1665$, $f=4.15659$ and $f=5.99865$. It is observed that there is a significant difference in the rate of occurrence between positive valued and negative valued extreme events at $f=3.1665$ whereas there are only negligible difference at $f=4.15659$ and $f=5.99865$. The IEI - PDF is calculated for $10^{9}$ time iterations.
	
	\section{Mechanism}
	\label{sec:5}
\begin{figure}[!ht]
		\centering
		\includegraphics[width=12cm]{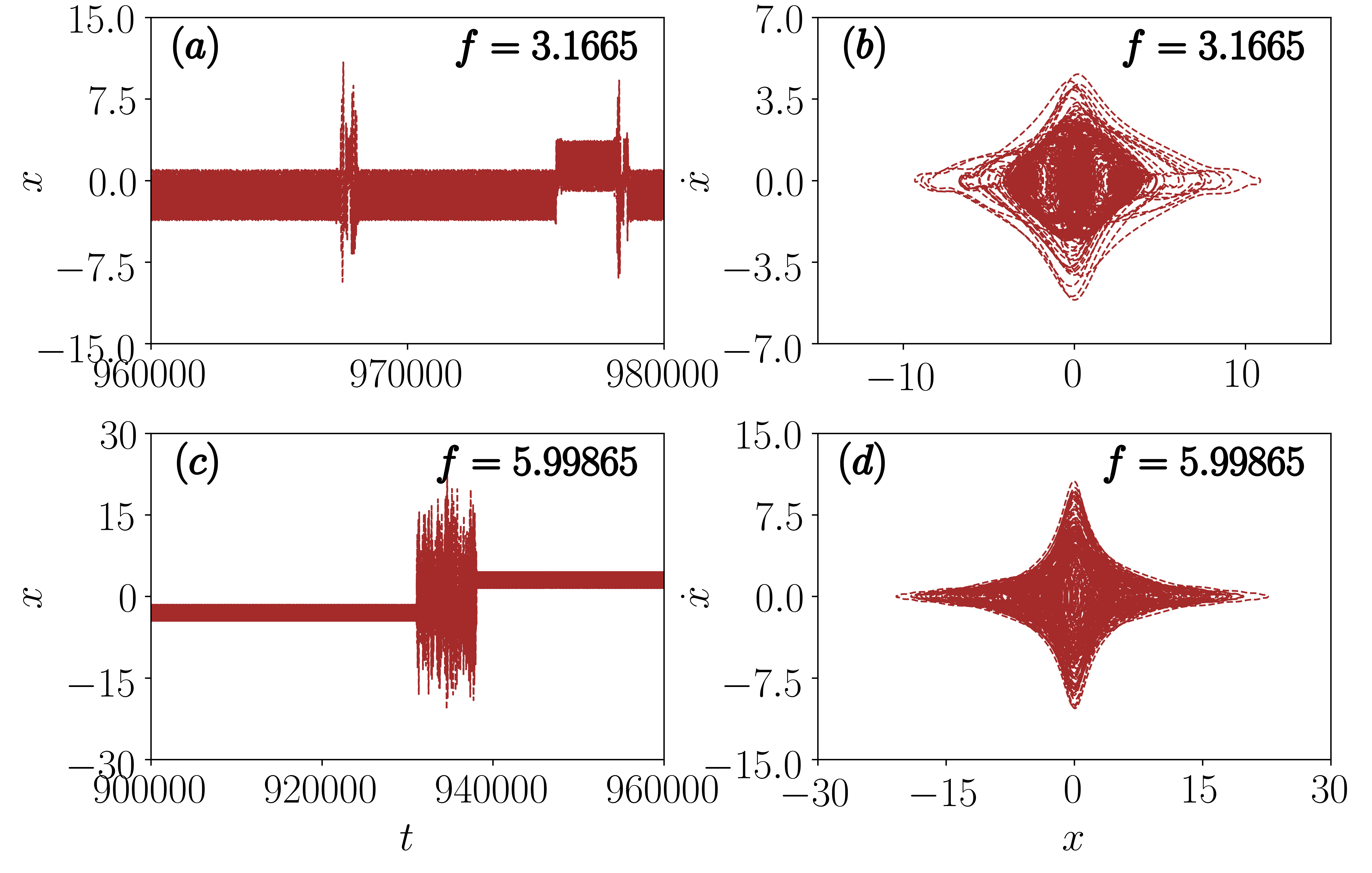}
		\caption{Enlarged time series and phase portraits showing mechanism under category $I$. Left panel is the enlarged time series and the right panel is the phase portrait}\label{mec1}
	\end{figure}

\begin{figure}[!ht]
		\centering
		\includegraphics[width=12cm]{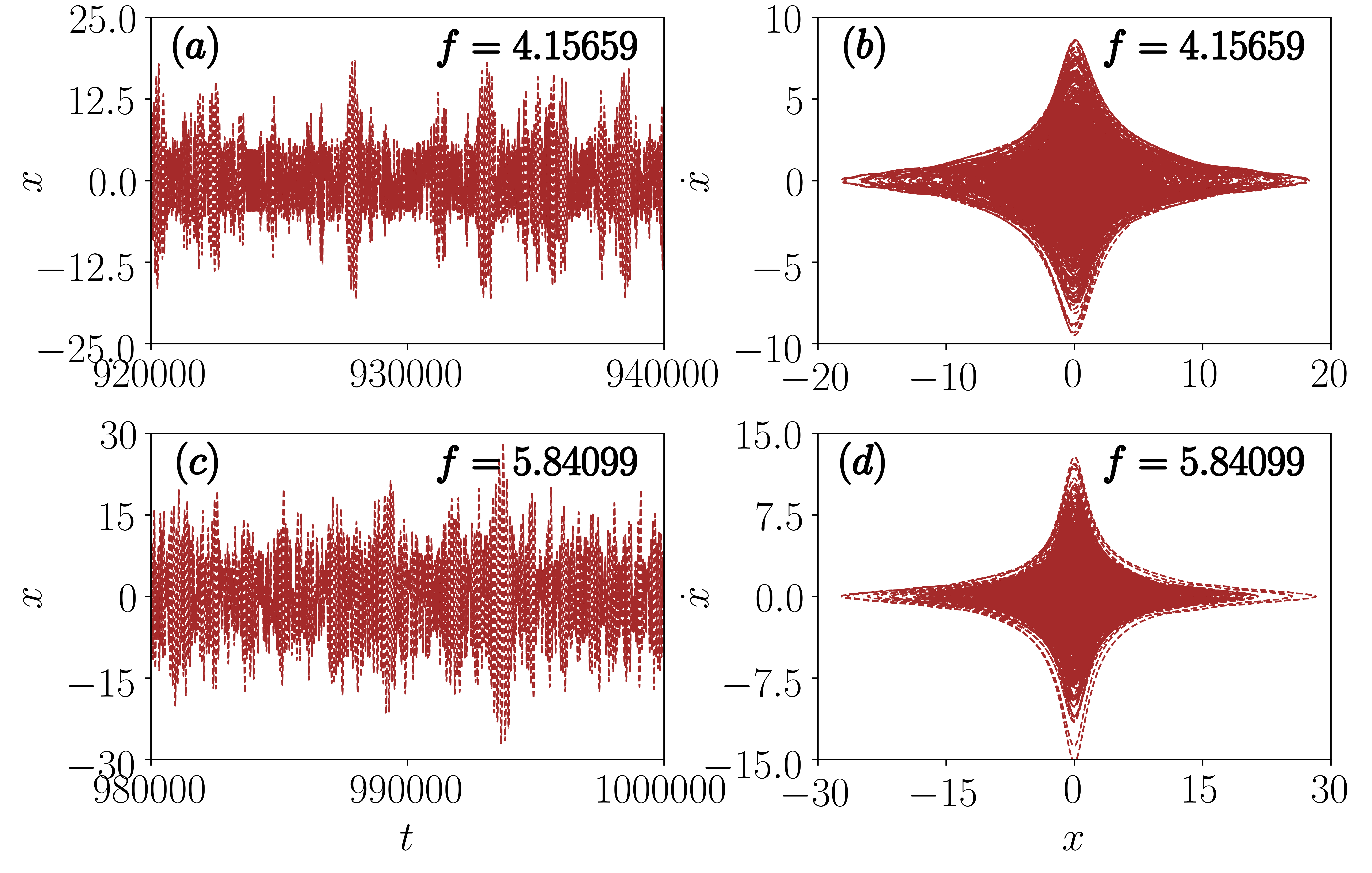}
		\caption{Enlarged time series and phase portraits showing mechanism under category $II$. Left panel is the enlarged time series and the right panel is the phase portrait.}\label{mec2}
	\end{figure}

	In this section, we analyze how the extreme events occur for different $f$ values mentioned in the previous section. To study this, we plot the time series and phase portrait for all the values at which symmetrically extreme events emerge. The time series and phase portrait which we displayed in Figs.~\ref{mec1} and \ref{mec2} are the enlarged portions (for a shorter time period) of the time series plots given in section \ref{sec:3}. The enlarged pictures represent the time domain in which very few times extreme events occur. From these figures we determine the way in which these extreme events originate. 
	
\begin{figure}[!ht]
		\centering
		\includegraphics[width=12cm]{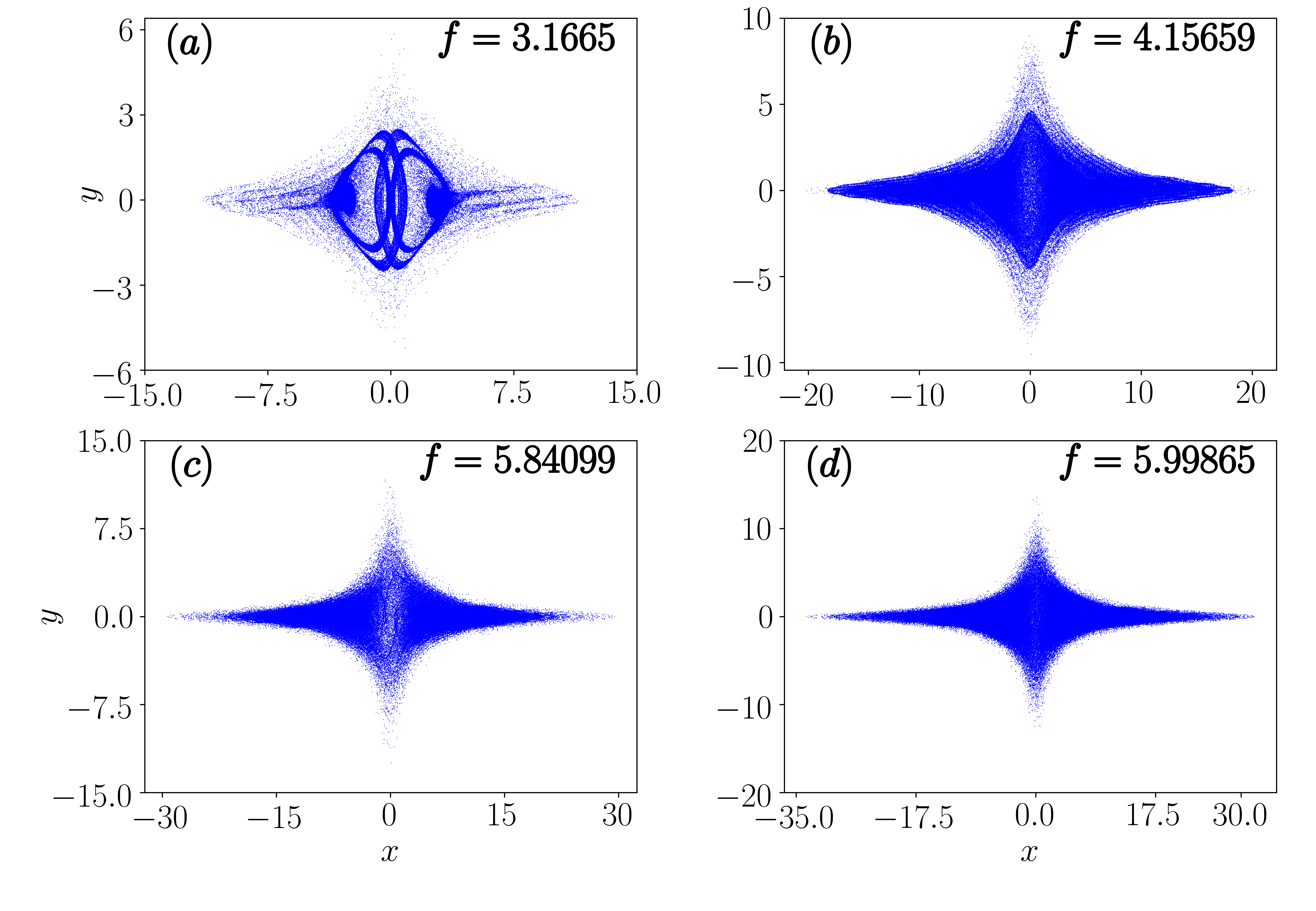}
		\caption{Poincar\'e cross section for a clocking time of $\frac{2\pi}{\omega}$ at four different points of Eq. (\ref{couppardri}). The parameter values are $\omega_0^2=0.25$, $\lambda=0.5$, $\alpha=0.2$, $\omega=1.0$.}
		\label{poincare}
	\end{figure}

	The Poincar\'e cross section for a clocking time of $\frac{2\pi}{\omega}$ near the points of interior crisis ($f=3.1665$ and $f=5.99865$) and near the points of tangent bifburcation ($4.15659$ and $5.84099$) of chaotic attractor is shown in Fig. \ref{poincare}.

	From Fig. \ref{poincare}a we can infer that the system exhibits a double band chaos. It can further be noticed that both the bands are symmetrically placed in phase space. At $f=3.1665$, the trajectories change then and there between the right and the left attractors. This occurs near the region of interior crisis where the sudden expansion of chaotic attractor is taking place. Similar behaviour can also be seen in Fig. \ref{poincare}d. This dynamics corresponds to the value of the parameter near interior crisis. The Poincar\'e cross section in Fig. \ref{poincare}b and \ref{poincare}c shows the behaviour  of the system near to the point of tangent bifurcation. Here, suddenly a large sized chaotic orbit appears. It is the reason why it looks like a single attractor. Once the chaotic attractor is emerged the amplitude of the chaotic attractor keeps on increasing continuously as $f$ increases. All the transitions occur symmetrically along the $x$ and $y$ axis.
	
	Based on our observation, the points at which the emergence of extreme events occur in the system (\ref{master}) can be classified into two categories. The first category (Category $I$) is due to interior crisis and is represented in Fig. \ref{mec1} while the second category (Category $II$) is due to tangent bifurcation and is represented in Fig. \ref{mec2}. 
	
	In Category $I$, extreme events occur near the point of interior crisis and occur as a burst deviation from the otherwise bounded chaotic orbit. In this case, the system exhibits bounded chaotic motion and in between intermittent bursting occurs to produce both large amplitude events and extreme events. That is, the system lies in one of the chaotic attractors for a very long time and when it tries to suddenly jump to the other attractor, the system's trajectory  travels across both the attractors for a period of time during which extreme events occur. This continues untill the trajectory moves completely into the domain of one of the attractors.  The time period of this burst is irregular and varies from region to region. In the second category, extreme events occur near the point of tangent bifurcation and occur amidst very large frequent amplitude oscillations. In comparison with the first category, the system exhibits large number of large amplitude oscillations and extreme events occur in between. This is because, before the bifurcation point, the system exhibits a $5t$ periodic orbit, and after the bifurcation point the system exhibits a chaotic motion. Now at the point of bifurcation (in our case $f=4.15659$ and $f=5.84099$) a new chaotic attractor is born and the system oscillates between periodic behaviour and chaotic behaviour resulting in frequent large amplitude events and extreme events indeed. The difference between the categories $I$ and $II$ is quite well observable from the time series and phase portraits described in the sub-figures (a)-(d) of Figs. (\ref{mec1}) and (\ref{mec2}) and from the Poincar\'e cross section shown in Fig.~\ref{poincare}. 
	
	Thus it is obvious that in both the cases, such extreme event occurs whenever a sudden expanded chaotic attractor appears. The difference between these two categories is, in Category $I$ chaotic attractor expands as a reason of interior crisis and in Category $II$, chaotic attractor emerges as a reason of tangent bifurcation. In Category $I$, instabilities inherent to the system causes such a large excursion to occur but the inherent chaotic nature of the system causes the trajectory to come back either to the attractor from where it originally started or to the other attractor. This is the reason for why phase portraits of category $I$ are sparse and two attractors are visible. In Category $II$, it is the nature of the point itself that causes extreme events to occur.  Hence in category $II$, the phase portraits are dense and looks like a single chaotic attractor. Similar behavioural difference can also be infered from the Poincar\'e cross section in Fig. (\ref{poincare}).
	
\begin{figure}[!ht]
		\centering
		\includegraphics[width=8.6cm]{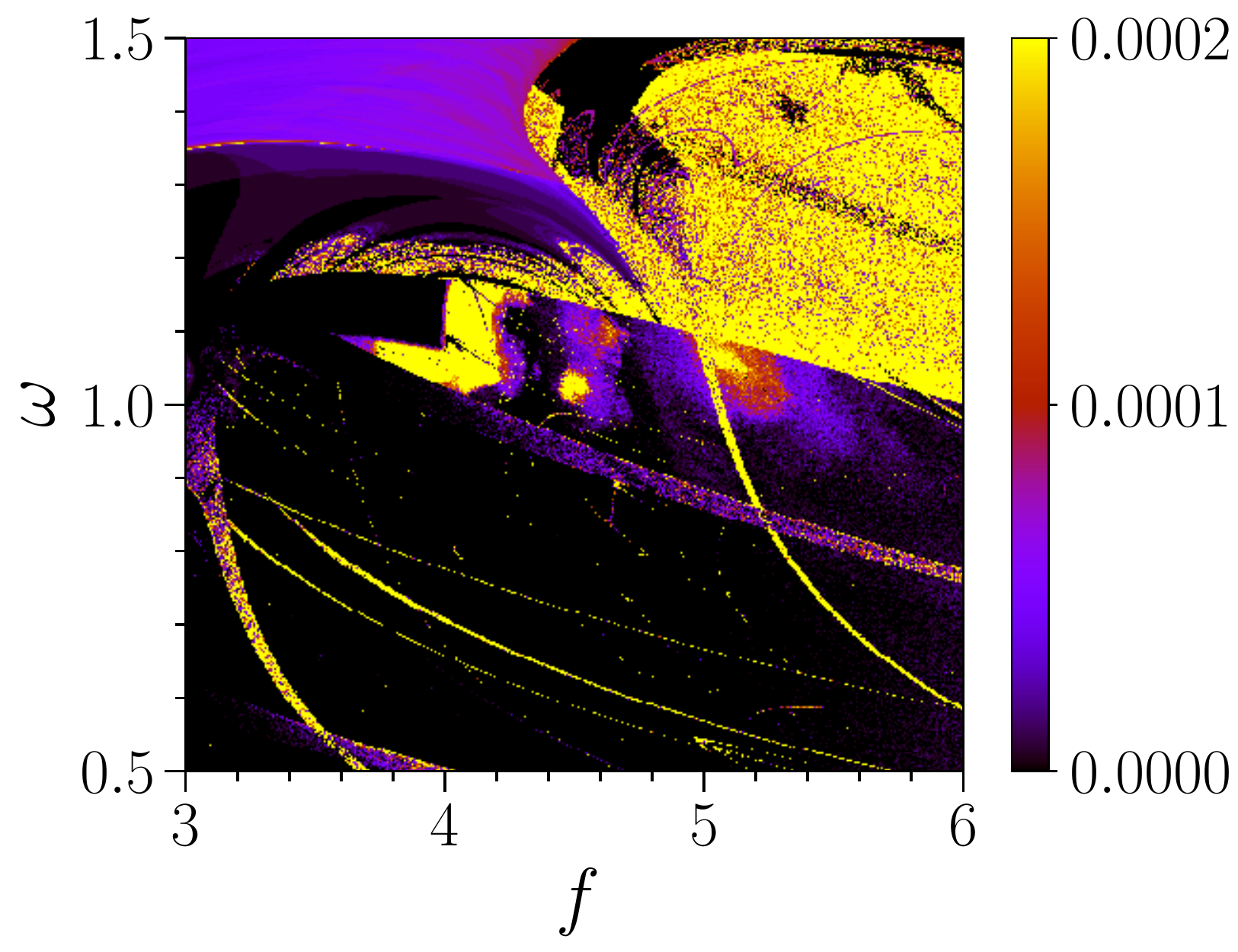}
		\caption{Two parameter diagrams in $f-\omega$ plane. Black spots represents regions with no extreme events and yellow spots (grey) represents the regions which have the  probability of occurrence of extreme events greater than or equal to $2\times10^{-4}$.} \label{2phase}
	\end{figure}

	
	The complete overview of the occurrence of extreme events in the two parameter ($f-\omega$) space is shown in Fig. \ref{2phase}.  Black spots represent regions with no extreme events and yellow spots (grey) represent the regions which have the  probability of occurrence of extreme events greater than or equal to $2\times10^{-4}$. It is clear from the two phase diagram that, for $\omega<1.0$ there are only sparse emergence of extreme events whereas, for $\omega>1.0$, there are more probability for the occurrence of the extreme events. From Fig~\ref{2phase}, it can be seen that extreme events with high probability occurs only for $f\ge4$. 
	
	Thus, as far as the damped and driven, velocity-dependent mechanical system (\ref{master}) is concerned, multiple and symmetrical extreme events occur as a result of interior crisis and tangent bifurcation.
	
	\section{Conclusion}	
	In this work we have reported the emergence of extreme events at multiple points in a damped and driven velocity-dependent system (\ref{master}). The emergence was found to occur symmetrically in both positive and negative spatial directions. We have also determined the mechanism by which extreme events occur in this system. It is found that the emergence of events can be classified into two categories. In category $I$, extreme events occur as bursts while in category $II$ extreme events occur amidst multiple large amplitude oscillations. In particular extreme events are found to occur at the points of interior crisis and tangent bifurcation. Since most of the real world systems are highly nonlinear, determining the emergence of extreme events in a highly nonlinear system such as the non-polynomial damped and driven velocity-dependent system considered in this work turns out to be very important. Further, our results highly help in the design of real-time mechanical and engineering rotating systems in order to avoid the occurrence of extreme events and to mitigate its disastrous effects.
	
	\section{Acknowledgement}
	The authors are thankful to Prof. P. Muruganandam, Department of Physics, Bharathidasan University, for his personal help in generating the two parameter diagram. 
	SS  thanks  the  Department  of  Science  and  Technology (DST), Government of India, for support through INSPIRE Fellowship (IF170319).  The work of AV forms a part of a research project sponsored by DST under the Grant No.EMR/2017/002813. The work of MS forms a part of a research project sponsored by Council of Scientific and Industrial Research (CSIR) under the Grant No. 03(1397)/17/EMR-II. 
	\\

\end{document}